# Community-Centered Resilience Enhancement of Urban Power and Gas Networks via Microgrid Partitioning, Mobile Energy Storage, and Data-Driven Risk Assessment

Arya Abdollahi
Polytechnic University of Bari, Bari, Italy

**Abstract:**
Urban energy systems face increasing challenges due to high penetration of renewable energy sources, extreme weather events, and other high-impact, low-probability disruptions. This project proposes a community-centered, open-access framework to enhance the resilience and reliability of urban power and gas networks by integrating microgrid partitioning, mobile energy storage deployment, and data-driven risk assessment. The approach involves converting passive distribution networks into active, self-healing microgrids using distributed energy resources and remotely controlled switches to enable flexible reconfiguration during normal and emergency operations. To address uncertainties from intermittent renewable generation and variable load, an adjustable interval optimization method combined with a column and constraint generation algorithm is developed, providing robust planning solutions without requiring probabilistic information. Additionally, a real-time online risk assessment tool is proposed, leveraging 25 multi-dimensional indices including load, grid status, resilient resources, emergency response, and meteorological factors to support operational decision-making during extreme events. The framework also optimizes the long-term sizing and allocation of mobile energy storage units while incorporating urban traffic data for effective routing during emergencies. Finally, a novel time-dependent resilience and reliability index is introduced to quantify system performance under diverse operating conditions. The proposed methodology aims to enable resilient, efficient, and adaptable urban energy networks capable of withstanding high-impact disruptions while maximizing operational and economic benefits.

**Introduction:**
This project initiative is to create a community-focused, open-access, and open-source tool for enhancing energy *and gas* resilience in urban areas. The outcome of this project is to design a tool to reconfigure and deploy mobile energy storage technologies strategically.

In the last several years, extensive research has been carried out to investigate the partitioning of distribution networks into interconnected $\mu$Gs, especially when dealing with faults and disruptions, which has been explored in various studies [1]-[5]. Specifically, authors [1] introduced a stochastic sectionalization method for distribution networks to enhance the system's resilience to infrequent, high-impact events. They achieved this through the implementation of a rolling horizon optimization strategy. In [3], the authors have explored $\mu$Gs as an alternative, in contrast to traditional methods of expansion planning for generation and transmission. In their research, they have divided the planning of microgrids into a two-level model. The primary level focuses on the overall planning of microgrids, while the secondary level deals with the operational scheduling of these μGs.

In [4], a recommended strategy involves a two-stage robust optimization method for determining the most suitable capacity and placement of multiple distributed energy resources within the μG. The primary objective is to minimize the overall long-term expenses while concurrently maximizing the profitability of the system operator. In [5], a μG type selection methodology has been presented in conjunction with distributed generation planning. This methodology aims to identify the optimal combination of hybrid AC/DC μGs to reduce the investment and operational expenses associated with DGs effectively. Additionally, it seeks to minimize the costs of purchasing energy from the upstream grid and the expenses associated with enhancing grid reliability.

An extensive analysis of the latest developments in μG planning problems underscores the presence of several essential deficiencies that must be adequately tackled. For instance, their primary emphasis has been on the interconnected operation mode of μGs (i.e., normal operational conditions), while the isolated operation mode (i.e., self-healing actions) of microgrids have not received thorough examination within the context of μG planning issues [2], [3], [5]. Furthermore, it is imperative to create novel metrics for assessing the reliability and resilience of μGs during islanding mode, as this important aspect has not yet been explored in previous research [6], [7]. Also, recent and ongoing investigations reveal numerous solutions. Various mathematical frameworks, including multi-objective modeling [8], bi-level modeling [9], and two-stage programming [10, 11], can be used to develop μG planning, depending on their complexity and objectives.

In evaluating distribution system risk, two key methodologies are mainly utilized: simulation techniques and analytical methods [12]. In distribution system risk assessment, simulation techniques, including sequential [13] and non-sequential [14] Monte Carlo simulations, use several probability models to determine the system states representing its components. Conversely, analytical methods [15] only work due to a limited set of system states with elevated probabilities. Nevertheless, techniques falling within both types require access to consistent topological and parameter data that pertains to a particular distribution network. These values can vary due to factors like meteorological conditions, system reconfiguration, and other similar considerations. Moreover, contemporary distribution networks encompass numerous uncertainties, including renewable energy sources, controlled loads, and electric vehicles, which are not comprehensively addressed within traditional methodologies.

Over the past few years, data-driven risk assessment approaches have garnered increasing attention. The authors in [16] delve into analyzing various elements, including supply disruptions, faulty equipment, unfavorable meteorological conditions, and other pertinent factors, in the context of evaluating distribution system risk. In [17], the assessment of operational reliability (or risk) combines big data technology with an artificial neural network. The data sources are classified into several distinct types, including financial, operational, and load-related data. An enhanced logistic regression pattern has been presented in [18] for predicting operational reliability to address the challenge of needing a substantial volume of data to train an ANN. This model incorporates numerous meteorological variables into its analysis. The research mentioned above primarily concentrates on the regular operation of distribution systems and isn't suitable for assessing the risks associated with distribution systems that demonstrate resilience during extreme events. This limitation arises from omitting crucial factors such as resilient resources and emergency response capabilities. The preference information is often assessed using simplified training models, and this assessment is commonly conducted through the application of the analytic hierarchy process and fuzzy comprehensive evaluation, as seen in references [19].

A recently introduced statistical risk assessment model evaluates potential grid disruptions, emphasizing guidance for renewable energy trading among prosumers. It utilizes the rare events logistic regression model for weather impact analysis and employs feeder taxonomy and hierarchical clustering to identify equipment faults tied to geographical features [20]. The authors in [21] present weather-driven analytics for spatial-temporal electricity generation forecasts and reliability evaluation. The proposed decision support tool enables operators to assess high-risk weather threats proactively. Highlighting the importance of power system resilience, the authors in [22] present a two-stage framework for assessing resilience against hurricanes. The paper introduces the conditional value at risk as a risk measure to evaluate the system's resilience against varying hurricane intensities.

An online spatial risk analysis using a severity risk index indicated in [23] to assess the impact of extreme events on system resilience integrates real-time monitoring, probabilistic risk, and scenario reduction to provide operators with a probabilistic assessment for effective resilience-based decision-making. A fast risk assessment method for distribution grids during extreme weather is given in [24], utilizing probabilistic graphical models. A dynamic risk assessment distribution system based on a comprehensive vulnerability rate model is introduced in [25]. The authors in [26] address high-impact, low-probability events in urban distribution networks, proposing a multi-stage resilience enhancement framework. The framework involves scenario-based planning, risk assessment, and plan amendments to reduce second-order impacts during ice disasters.

Huang and colleagues in [27] introduce a comprehensive approach that combines topology switching and proactive load shedding to enhance the system's ability to respond effectively to resilience challenges. An approach presented in [28] to improve the efficiency for RES, stationary battery storage, and electric vehicle parking within distribution systems (DS) is aimed at optimizing planning for both normal and resilient operations. The authors in [29] initiated an organized methodology for segmenting sections of a distribution system into microgrids featuring flexible boundaries to improve the resilience index for low-probability, high-impact events. Shi and fellow researchers [30] have formulated a strategy to enhance the resilience index during preventive and post-disaster phases in response to severe meteorological conditions. This includes integrating mobile energy sources dispatched to diverse positions for supporting μGs.

An approach was proposed in [31] to assess the resilient index in multi-μGs, highlighting the capacity of mobile energy units to alleviate the negative impacts of natural disasters and enhance the overall resilience of the system. The authors aim to strengthen the resilience of the distribution network facing earthquakes by addressing roadway congestion and outage time as key challenges in the initial hours following the event, leveraging the capabilities of a mobile energy storage unit [32]. The authors in [33] create a tri-objective model to address optimal sizing and pre-positioning challenges driven by resilience for mobile energy storage units in interconnected microgrids. A dynamic approach is suggested to strategically develop islands within an islanded microgrid, aiming to maximize the supply of demand over the planning time horizon. Additionally, the study incorporates mobile energy storage units mounted on trucks [34]. A newly devised restoration mechanism in a distribution network that focuses on the routing and scheduling mobile energy storage systems combined with unpredictable RES is modeled in [35] to attain a resilient system for addressing high-impact, low-probability events. In [36], a two-stage stochastic model is introduced to maximize the utilization of mobile energy storage units. In the initial stage, the focus is on defining the planning decisions related to the positions and capacities of mobile storage units. Subsequently, in the next stage, the assessment of operational costs for the active distribution network considers normal, severe, and extreme conditions.

**Gaps:**

- Gap 1: There is not that much focus on resilient strategies for future power grids with high penetration rates of renewable energy sources, which hinders the development of practical solutions for addressing high-impact and low-probability events, including hurricanes, floods, ice storms, cyberattacks, internet system attacks, etc.

- Gap 2: Despite the increasing recognition of severe uncertainty in decision-making contexts, existing methodologies often fail to address this challenge comprehensively. A noticeable research gap exists in the pursuit of methods that combine creativity and robustness to enhance the realism of results in such uncertain environments.

- Gap 3: There is no model to optimize the long-term sizing and allocation of mobile storage devices and short-term operation for resilience enhancement that considers the optimal routing of the storage. Also, despite the growing interest in mobile energy storage systems and their potential to optimize urban energy distribution, there is a significant research gap in developing modeling approaches that effectively integrate real-time urban traffic data to achieve accurate and optimized routing for these systems. Existing studies primarily focus on the energy storage systems' technical aspects, neglecting the crucial interplay between energy storage, traffic dynamics, and urban infrastructure.

- Gap 4: Existing resilience assessments often rely on reliability metrics, which may not fully capture system performance under diverse operating conditions. Developing new, more accurate resilience indices is an essential research challenge. Therefore, there is a need to develop an integrated resiliency-reliability index for energy systems.

**Objectives**:

- **Objective 1:** Systematically analyze and propose resilient strategies specific to distribution systems with significant renewable energy integration, aiming to enhance their resilience to high-impact and low-probability events based on spatiotemporal.

- **Objective 2:** Develop a data-driven risk-assessment framework to address severe uncertainty and risk associated with high-impact and low-probability events based on advanced data analytics, machine learning, and artificial intelligence to enhance the accuracy of predictions.

- **Objective 3:** T1: Develop a framework for optimizing the routing of mobile storage devices while simultaneously integrating long-term design strategies, to enhance operational efficiency and align routing decisions with broader, forward-looking organizational goals and requirements. T2: The objective is also to create a modeling approach for mobile energy storage systems that incorporates urban traffic considerations and aims to achieve accurate and optimized routing for these systems.

- **Objective 4:** Propose a novel time-dependent index to accurately measure the resilience of integrated energy systems when faced with catastrophic events.

**Tasks**:
- **Task 1**: Distribution system operators (DSOs) still have challenging responsibilities when it comes to the optimal operation and plan of μGs, which is why a lot of recent study has been devoted to addressing these critical concerns. To provide resilience and reliability requirements, a critical task is to extend an optimal model for effectively dividing distribution systems into a group of self-healing μGs. To improve resilience and self-healing, μGs can physically be linked to each other at various locations using remotely controlled switches (RCS) to transfer power between themselves (i.e., spatial) during emergency and normal conditions (i.e., temporal). The concept of self-healing in this context can be understood as the inherent capability of the power system to effectively employ corrective and preventive measures in response to faults, thereby mitigating the influence of unforeseen disruptions on the system's operational performance.

   This proposed project will develop an effective methodology for partitioning distribution systems into resilient self-healing μGs, considering uncertainties. Create a compelling and adaptable framework for dividing distribution systems into a cluster of μGs that are resilient, facilitating their management under normal and self-healing conditions. The research initiative will encompass two fundamental phases for the conversion of large-scale distribution systems into interconnected, supply-sufficient μGs:
   i) Convert passive distribution networks into active ones by integrating different RES and mobile energy storage, transforming them into dynamic and resilient entities.

ii) Establish the distribution network as networked μGs by defining electrical and geographical limitations through the allocation of RCS, ensuring efficient and well-defined microgrid clusters.

We aim to model the parallel evolution of diverse renewable energy sources and delineate the electrical and geographical parameters of μGs. Our research team will incorporate reliability and resilience considerations into the planning phase, introducing innovative metrics for evaluating μG reliability in interconnected and isolated modes. Addressing the inherent uncertainty of the challenge, we will employ adjustable interval optimization and utilize a column and constraint generation (C&CG) approach for problem resolution. The research team will execute this task through two sub-tasks to guarantee sufficient data for our proposed methodology's development.

**Sub-task 1: Framework/schematic data collection**
Michigan, DTE, Consumers Energy, Equipment, Installation, Feature of data (time, location, type of outage, solar, wind speed, etc.)
We will utilize the electric distribution network in Detroit to apply our proposed approach. The problem at hand will be executed using Matlab and GAMS environments. The technical parameters and levelized cost of energy associated with Distributed Energy Resources (DER) units have been sourced from existing references. In our project, we will assume a planning horizon of five years. The Detroit electric distribution company will provide the load profiles for μGs on an annual basis, accounting for all changes every year. Additionally, our project necessitates the hourly electricity prices from the retail market, obtainable from the Michigan power market, for use in simulations. The DER units, identified as the backbone of μGs, are tasked with meeting the local demand of μGs. As a result, these resources must possess sufficient capacity to fulfill the critical loads of μGs through islanding mode. It's important to emphasize that the suggested problem is static, meaning all investments will be completed in the initial year. Within the specified distribution network, μG is equipped with RCS to minimize load curtailment through cascading outages. It is crucial to highlight that the cumulative capacity of DER placed at every μG must exceed its critical loads.

**Sub-task 2: Model development**
The main objective of partitioning the distribution system into autonomous, self-healing microgrids is to maximize the distribution system operator's benefit during normal operation while minimizing the need for load shedding when operating independently in island mode. In fact, in normal operation mode, the distribution system operator aims to reduce its costs by optimally dispatching its distributed generation resources and exchanging power between the microgrids while following a fault in the distribution network; microgrids enter islanding mode to maintain system reliability as much as possible. The main goal in the island mode is to maximize the reliability of the system instead of gaining economic benefits, using some different self-healing measures such as reconfiguration, rescheduling of distributed generations, and load shedding.

Our team has chosen to tackle the problem by formulating this problem as a linear and convex model, primarily due to the advantages these models offer in terms of convergence and optimality. Our objective centers around maximizing profits, which can be expressed as the difference between the revenue generated and the costs incurred over the planning horizon. Revenue is generated through the sale of power to customers and the power market, while costs encompass various economic factors related to both the initial investment and ongoing operation of RES and RCS within μGs. It is important to note that we have unified the investment and operational costs associated with RES and RCS into a single, cost-based objective function.

Considerable uncertainties, such as the intermittent nature of renewable energy generation, present considerable challenges when maximizing the efficient utilization of renewable resources. To explore how these uncertain factors affect the issue, our team will introduce an interval optimization tool that leverages the C&CG framework. In interval optimization, we will consider the uncertain parameter to fluctuate within a defined range of minimum and maximum values to determine the feasible computational time required to find both the lower and upper boundaries of the output results. To explain further, interval optimization offers a method to establish highly reliable solution bounds with complete confidence for modeling objectives. This is achieved by enabling the computation of interval bounds for precise integral values through branch-and-bound techniques. In contrast to stochastic scenario-based programming, where understanding the probability density function (PDF) of uncertain parameters is essential and introduces considerable computational complexity to the problem, interval optimization eliminates the need for knowledge of the PDF of parameters in the decision-making process.

**Adjustable Interval Optimization**

To employ the interval optimization method for the given model, we will express the deterministic planning model for μGs in a standard form, presenting the relevant inequality and equality constraints in their standard equations (1-3).

$$\underset{DV}{Max} \ F = profit\left[\Psi, \xi\right] \tag{1}$$

$$X(M, N, \Psi) \leq 0 \tag{2}$$

$$Y(M, N, \Psi) = 0 \tag{3}$$

Supposing $\xi$ represents an uncertain parameter with lower and upper limits $\xi$min and $\xi$max, the objective function's minimum and maximum amounts can be computed using equations (4) and (5) as opposed to relying on the expected value.

$$F^-(\Psi) = \min_{\xi} F(\Psi) \tag{4}$$

$$F^+(\Psi) = \max_{\xi} F(\Psi) \tag{5}$$

Following this modification, the main objective function is replaced with the interval objective function, which optimizes its bounds inside the designated input intervals. Stated differently, interval optimization does not only consider the worst-case situation; it also tries to reduce the objective function's range. To achieve this goal, it is necessary to simultaneously optimize the average goal ($F_{div}$ ($\Psi$)) and its variants ($F_{ave}$ ($\Psi$)), as shown in equation (6), until the μGs show improved resilience against uncertainty. The goal function's average and variation are described in detail in expressions (7) and (8).

$$F(\Psi) = Max\,(F^{avg}(\Psi)\,|\,Min\;F^{div}(\Psi)) \tag{6}$$

$$F^{avg}(\Psi) = \frac{F^+(\Psi) + F^-(\Psi)}{2} \tag{7}$$

$$F^{div}(\Psi) = \frac{F^+(\Psi) - F^-(\Psi)}{2} \tag{8}$$

*Proposed C&CG algorithm*

Evidently, the proposed model poses a challenge as it involves a min-max-min problem, which is not directly amenable to standard commercial solvers. To tackle this complexity, we propose employing a C&CG framework. This approach dissects the intricate problem into solvable sub problems, including a master problem and a set of slave problems. By doing so, the original problem becomes solvable using widely available mixed-integer linear programming (MILP) solvers. Utilizing the suggested C&CG framework, the master problem is defined as follows:

$$\min_{f \in F} A_0^T f + \xi \tag{9}$$

$$\text{s.t. } \xi \geq B_0^T u + C_0^T z \tag{10}$$

$$A_1^T f + B_1^T u + C_1^T z = q_1 \tag{11}$$

$$A_1^T f + B_1^T u + C_1^T z \leq q_2 \tag{12}$$

The initial problem is evidently relaxed in the master problem. Confirm that the solution $f$ corresponds to the master problem. Additionally, describe the sub problem using the following definition:

$$\max_{u \in U} \min_{z \in Z} B_0^T u + C_0^T z \tag{13}$$

$$\text{s.t. } A_1^T \hat{f} + B_1^T u + C_1^T z = q_1 \tag{14}$$

$$A_1^T \hat{f} + B_1^T u + C_1^T z \leq q_2 \tag{15}$$

For any non-optimal solution (*f*), the sub-problem's objective value is larger than the main problem's true optimal solutions. Consequently, the sub-problem yields an upper bound for the problem. Still, the sub-problem is a bi-level problem that is challenging to resolve. Through by applying Karush-Kuhn-Tucker (KKT) optimality conditions, we will reformulate constraints (13)-(15) as complementary constraints (16)-(18).

$$A_1^T \hat{f} + B_1^T u + C_1^T z = q_1 \tag{16}$$

$$C_0^T + \varphi^T C_1 + \lambda^T C_2 = 0 \tag{17}$$

$$0 \le q_2 - C_2 z - A_2 \hat{f} - B_2 u \perp \lambda \ge 0 \tag{18}$$

In the expressions (13)–(15), $\varphi$ and $\lambda$ serve as the dual variables. To ensure convexity, the KKT conditions can be transformed into a linear form using the Big-M approach (19, 20).

$$0 \le q_2 - C_2 z - A_2 \hat{f} - B_2 u \le M\theta \tag{19}$$

$$0 \le \lambda \le M \cdot (1-\theta) \tag{20}$$

An evaluation of the proposed approach has been conducted: The model can effectively select a suitable mix of different RES portfolios to address the system's economic and reliability requirements. This approach aims to minimize the overall planning costs while facilitating the smooth transition to islanding when needed. Enhancing the variety of RESs in use enhances the resilience and dependability of microgrids when dealing with various short-term operational scenarios.

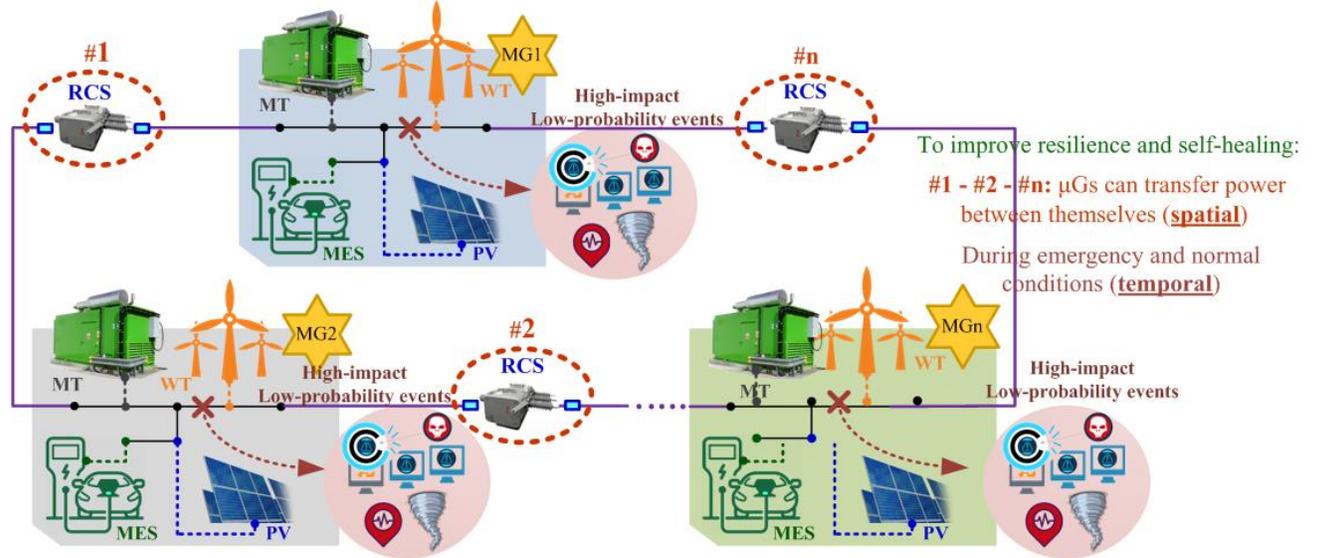

● **Task 2:** A practical online risk assessment tool is crucial to deliver warning data for distribution network resilience promptly. However, conventional analytical risk assessment methods are subject to known network information, while data-driven methods rarely incorporate resilient resources into the risk assessment procedures, limiting their accuracy when applied to high-impact and low-probability events. Conventional methods used in distribution system risk assessment can be divided into two categories: simulation and analytical methods. The simulation methods, like Monte Carlo simulation, evaluate the system states sampled by the probability models of system components. While the analytical methods are only based on a limited set of system states with higher probabilities, the methods in both categories need the known topology and parameter information of a particular distribution system, which is changeable due to weather conditions and reconfiguration.

Our team will develop an online data-driven risk assessment method that is adaptive for resilient distribution systems. Several basic operational indexes from practical experience are considered to reflect the system risk, and entropy weights and gray correlation degrees characterize the complicated relationship between the indexes and risk.

**Sub-task 1: Fundamental indices for online risk assessment**

The initial step involves filtering and identifying the primary risk factors to facilitate online risk assessment of distribution systems in the face of extreme events. We propose an organized index system for assessing risks, categorizing the indicators of resilient distribution systems into different dimensions, including electric load, power equipment and network, resources for resilience, emergency management, and geological and meteorological factors. Considering practical requirements and the viability of conducting online risk assessments, we select 25 resilient indicators, divided into 5 categories as outlined in Table 1, to serve as the foundational metrics for risk assessment.

| Table 1. Indexes for online risk assessment | | |
|---|---|---|
| No. | Category | Index |
| 1 | Load | Loss of load probability |
| 2 | | Anticipated outage duration |
| 3 | | Expected proportion of system load loss |
| 4 | | Predicted maximum load loss ratio |
| 5 | | Current load-shedding ratio |
| 6 | Grid | Mean lines' utilization rate |
| 7 | | Peak lines' utilization rate |
| 8 | | Typical transformers' usage rate |
| 9 | | Maximum transformers' usage rate |
| 10 | Resilient resources | Solar electricity generation prediction |
| 11 | | Wind energy generation prediction |
| 12 | | Real-time energy storage device capacity |
| 13 | | State of charge for energy storage devices |
| 14 | | Controllable load capacity |
| 15 | | Controllable load adjustment rate |
| 16 | | Live capacity of diesel power units |
| 17 | | Anticipated operational duration of diesel generators |
| 18 | Emergency response and repair resources | Count of mobile power generation units |
| 19 | | Expected operational duration of mobile power generation units |
| 20 | | Quantity of maintenance team personnel |
| 21 | | Inventory of repair materials |
| 22 | | Effectiveness of on-site repair resources |
| 23 | | Significance of the existing damaged equipment |
| 24 | Meteorology | Meteorological disaster alert threshold |
| 25 | | Current weather conditions intensity |

The real-time data for the 25 indices listed in Table 1 can be sourced from various sources in practical scenarios. Predictions for anticipated load details are generated through an online reliability assessment tool. Present load and grid data are derived from distant monitoring instruments. Information about resilient resources is both furnished and predicted by a distributed generation management system or manually inputted. As for data regarding emergency responses and repair resources, it is supplied by the emergency management unit within the distribution system operator. Ultimately, meteorological data is procured from the National Weather Service. Figure 1 illustrates the diverse origins of real-time data.

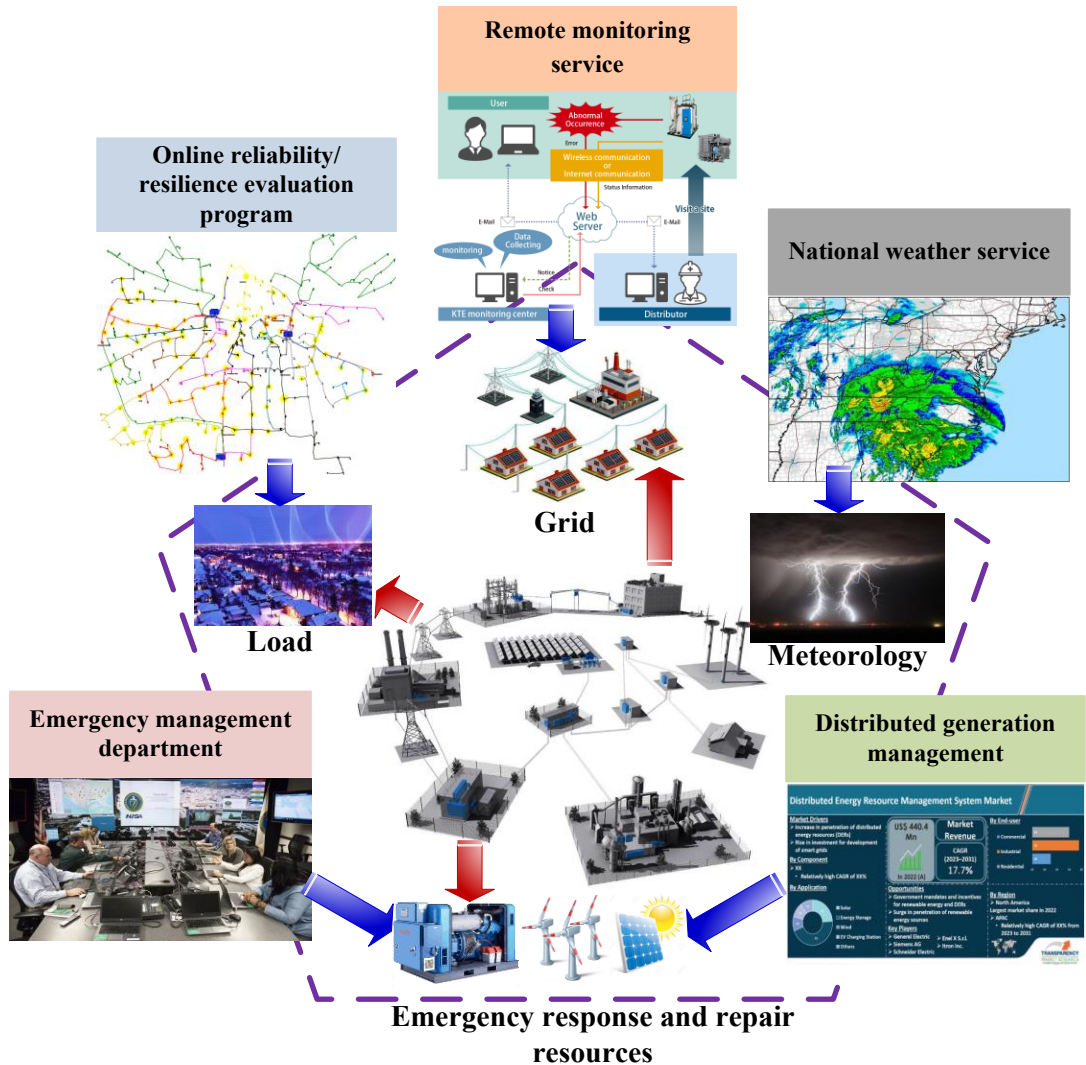

**Sub-task 2: Data collection**

To assess the effectiveness of our proposed approach on the Detroit electric distribution system, it is essential to initiate a data generation process before conducting the online risk assessment. For illustrative purposes, we have chosen to focus on the risk associated with hurricane weather. Still, it's worth noting that our method can be readily extended to address various other extreme events. The time-series load data utilized in this study is derived from the standard daily load profile of the Detroit area. The foundational principles underlying the creation of the historical time-series data, as outlined in Table I, are as follows:

1) The data collection involves generating short-term load data, designated as indices 1 to 5. These data points will be produced using the Failure Mode and effect analysis (FMEA) methodology within this project's scope. Our approach comprehensively evaluates N-1, N-2, and N-3 branch failure scenarios while considering the dynamic impact of real-time wind speed fluctuations on branch failure rates and repair times. To facilitate the analysis, we will simplify the assumption that the correlations between wind speed and failure rate, as well as wind speed and repair time, conform to predefined patterns.

2) We will employ a structured data collection approach in our pursuit of essential operational and resource data, specifically for indices 6–9 and 12-23. Our methodology commences with a systematic sequential sampling of branch states, distinguishing between normal and failure states. Subsequently, we will employ FMEA to capture grid operational data related to indices 6–9. Data collection for index 23 is straightforward and readily accessible. The values associated with resilient resources in indices 12, 14, 16, and 18 are contingent upon facility availability, which we will model using Bernoulli distributions. Initial data for indices 13, 15, 17, and 19 is randomly generated during outage scenarios, and subsequent data, including that for index 5, is acquired through a basic dispatch simulation of distribution system resources.
As for indices 20–22, data generation is based on the assumption of uniform distributions within their specified ranges.

3) In the context of weather-related data spanning indices 10-11 and 24-25, our data collection approach is as follows: To acquire renewable output forecast data related to indices 10-11, we will transform wind speed and solar irradiance data into power data using characteristic functions.
Regarding meteorological data associated with indices 24 and 25, explicitly concerning historical hurricane events, we will access this valuable information from the Detroit Electric Power Company.

- **Task 3:** Electrochemical energy storage (ES) units have been real-world-tested as an efficient backup resource that enhances the resilience of distribution systems. However, using these units for resilience is insufficient to justify their economic installation; therefore, these units are often installed in locations where they generate the most significant economic value during normal operations. Motivated by the recent progress in mobile ES technologies, like ES units that can be moved using public transportation routes, this objective will use this spatial flexibility to overcome the gap between the economically optimal locations during normal operations and the locations where extra backup capacity is necessary during disasters.

**Sub-task 1: Operation of mobile energy storage system**

**Normal operation:** We will examine the standard procedures where the distribution system operator employs mobile energy storage units to engage in spatiotemporal energy arbitrage during regular operations. Despite the potential mobility of energy storage in this scenario, these assets remain immobile during typical operations, aligning with the established practices of conventional utility companies.

**Emergency operation:** We will develop a resilience response strategy for the distribution system operator (DSO) to mitigate the impact and socioeconomic losses when a natural disaster is anticipated. Our proposed resilience response strategy suggests that the DSO can harness the flexibility of mobile energy storage units and other resilience technologies, such as topology switching and the formation of microgrids. Therefore, mobile energy storage units can be relocated from their fixed locations to address the disaster, making emergency operational decisions dependent on fixed siting decisions. To implement this, we will employ topology switching and microgrid forming to optimize the routing and deployment of mobile energy storage units.

In our laboratory, we intend to develop a planning model to optimize the investment costs associated with mobile energy storage units. This model will consider both normal and emergency operations, as discussed. The proposed planning model is outlined in Figure 2, illustrating the connections between planning and operation decisions and the decisions made during normal and emergency operations. The objective of this optimization is to allocate the mobile energy storage units so that these units are operated as stationary resources during normal operations and can be transported to other locations, or among multiple other locations, in case of natural disasters.

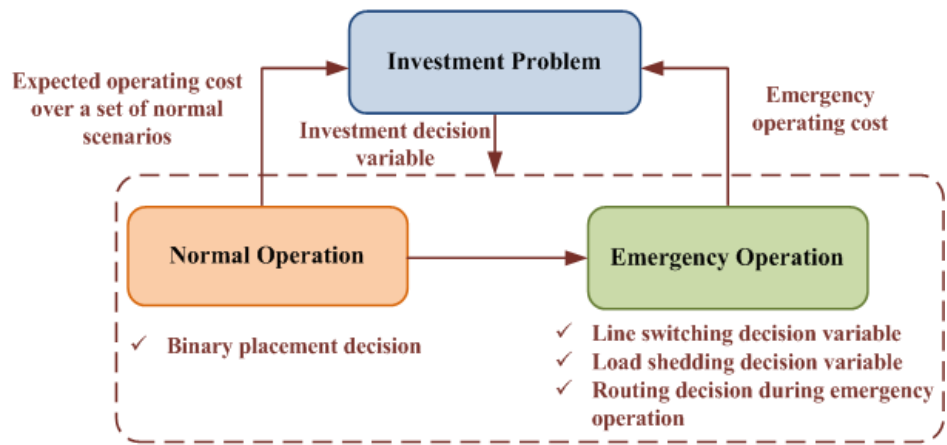

With this purpose, we suggest a two-stage optimization model that will optimize investments in mobile ES units in the first stage and can re-route the installed mobile ES units in the second stage to form dynamic microgrids and avoid the expected load shedding caused by disasters. This optimization will achieve the trade-off between the economic value of mobile ES units during normal operations and their ability to enhance distribution system resilience in the case of natural disasters.

During emergencies, ES units can travel among buses and form dynamic microgrids by using their spatial flexibility and optimizing the boundaries and centroids of microgrids. The optimization model will also formulate the transit delay of mobile ES units and switching decisions of distribution lines and loads. We suggest utilizing the progressive hedging algorithm to solve this optimization model, and the off-the-shelf solvers cannot efficiently solve this mixed-integer second-order cone programming problem with binary variables.

- **Task 4:** *First Stage - Pre-Disaster Preparedness*: During this stage, the operator endeavors to implement pre-disaster measures to optimize the system's readiness. This involves energy storage in designated facilities and enhancing the grid's resilience to potential configuration changes before the occurrence of any disruptive event. The mobile energy storage and natural gas storage states that should be optimized are displayed in the first part of OF (1). Furthermore, the second part of OF (1) features a graph-based metric that determines how resistant the grid topology is to configuration changes. Minimizing this metric is imperative to enhance the network's resilience against changes in its topology. We will employed the weight sum method to merge these two dimensions that vary in size and lack congruence into a singular objective. The first step is to equalize the dimensions of all parts by dividing each part by its base value.

$$\underset{H\&N}{\text{Max }} F^{H\&N} = \frac{1}{f_1^{\text{base}}} \sum_{t=1}^{T_c} \left\{ \sum_{e=1}^{N_e} \text{SoC}_{et}^{\text{MES}} + \sum_{g=1}^{N_g} \text{SoC}_{gt}^{\text{NG}} \right\} - \frac{1}{f_2^{\text{base}}} (\frac{2}{N-1}) \times \text{Trace}(L^+) \qquad (21)$$

*Second Stage - Post-Disaster Management*: We will introduce a novel time-dependent resilience index (RI), considering network vulnerability and weather hazards in light of evolving weather patterns and associated meteorological predictions (22). Acknowledging the uncertainties surrounding the impact and timing of extreme events, multiple scenarios will be considered, and decision-makers will seek to optimize for worst-case scenarios. Following the event, the operator prepares for the restoration process to return the system to its normal state swiftly. To facilitate this, a standardized metric for the restoration phase will be devised. The analysis of the system's resilience encompasses the entire spectrum from the initiation of degradation to the conclusion of the restoration process. In order to compute the anticipated Resilience Index (RI), it is essential to initially assess the system's susceptibility and deterioration in response to the envisioned event, as outlined in equations (23) to (24). It is important to note that these equations have been normalized to provide additional insights into the temporal dynamics of the system. A value of 0 signifies a fully resilient system, while a value of 1 indicates complete degradation.

Furthermore, following the conclusion of the event, the operator prepares to initiate the restoration process to swiftly return the system to its normal operating state. To facilitate this, a standardized measure for the restoration phase is outlined in equation (25). This metric is designed to attain a value of 1 for a system fully capable of restoration and 0 otherwise. It is imperative to analyze the system's resilience process comprehensively, spanning from the commencement of degradation to the completion of restoration. Therefore, equation (26) encapsulates the Resilience Index (RI) for a specific performance, with values ranging between 0 and 1, where elevated values signify greater resilience.

$$\underset{W\&S}{\text{Max}} \ F^{W\&S} = \sum_{\omega=1}^{N_\omega} \pi_\omega (RI_\omega + SI_\omega) \tag{22}$$

$$VI = \frac{M_0 - M_{pe}}{M_0} \tag{23}$$

$$DI = \int_{t_d}^{t_{pe}} (M_0 - M(t))dt \ / \ M_0(t_{pe} - t_d) \tag{24}$$

$$SI = \int_{t_d}^{t_{pr}} (M(t) - M_{pe})dt \ / \ M_0 - M_{pe}(t_{pr} - t_r) \tag{25}$$

$$RI = \int_{t_d}^{t_{pr}} M(t)dt \ / \ M_0(t_{pr} - t_d) \tag{26}$$

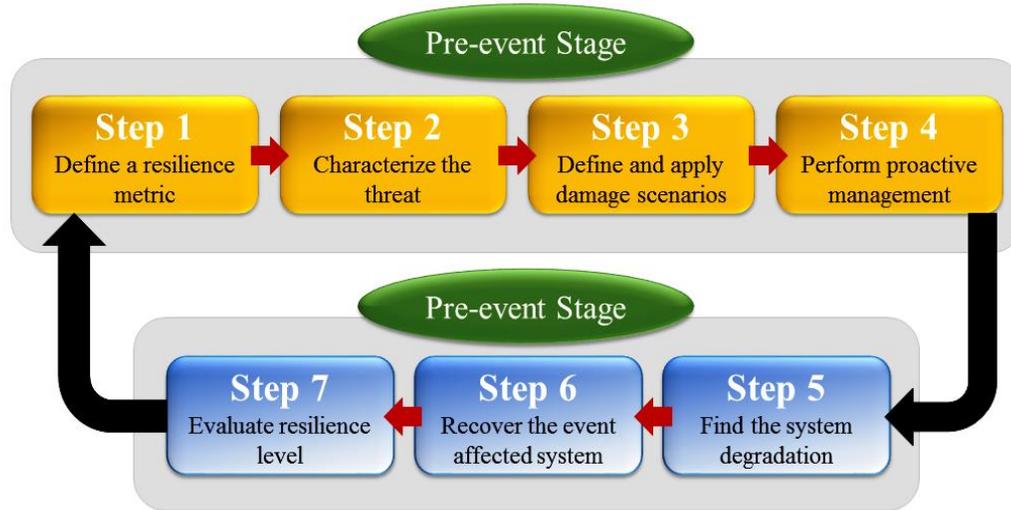

## References

[1] S. C. Khavar and A. Abdolahi, "Techno-Economic Dispatch of Distributed Energy Resources for Optimal Grid-Connected Operation of a Microgrid," in *2022 30th International Conference on Electrical Engineering (ICEE)*, May 17, 2022, pp. 507–511.

[2] R. R. Nejad and W. Sun, "Enhancing active distribution systems resilience by fully distributed self-healing strategy," *IEEE Transactions on Smart Grid,* vol. 13, pp. 1023-1034, 2021.

[3] A. Abdolahi, J. Salehi, F. Samadi Gazijahani, and A. Safari, "Probabilistic multi-objective arbitrage of dispersed energy storage systems for optimal congestion management of active distribution networks including solar/wind/CHP hybrid energy system," *Journal of Renewable and Sustainable Energy*, vol. 10, no. 4, Jul. 2018.

[4] S. M. R. H. Shawon, X. Liang, and M. Janbakhsh, "Optimal Placement of Distributed Generation Units for Microgrid Planning in Distribution Networks," *IEEE Transactions on Industry Applications,* 2023.

[5] A. Gagangras, S. D. Manshadi, and A. Farokhi Soofi, "Zero-Carbon AC/DC Microgrid Planning by Leveraging Vehicle-to-Grid Technologies," *Energies,* vol. 16, p. 6446, 2023.

[6] M. Abdel-Basset, A. Gamal, R. K. Chakrabortty, and M. J. Ryan, "Evaluation approach for sustainable renewable energy systems under uncertain environment: A case study," *Renewable energy,* vol. 168, pp. 1073-1095, 2021.

[7] M. Majidi, B. Mohammadi-Ivatloo, and A. Soroudi, "Application of information gap decision theory in practical energy problems: A comprehensive review," *Applied Energy,* vol. 249, pp. 157-165, 2019.

[8] Y. Zhu, G. Li, Y. Guo, D. Li, and N. Bohlooli, "Modeling Optimal Energy Exchange Operation of Microgrids Considering Renewable Energy Resources, Risk-based Strategies, and Reliability Aspect Using Multi-objective Adolescent Identity Search Algorithm," *Sustainable Cities and Society,* vol. 91, p. 104380, 2023.

[9] B. Huang, T. Zhao, M. Yue, and J. Wang, "Bi-Level Adaptive Storage Expansion Strategy for Microgrids Using Deep Reinforcement Learning," *IEEE Transactions on Smart Grid,* 2023.

[10] M. Nasir, A. R. Jordehi, M. Tostado-Véliz, S. A. Mansouri, E. R. Sanseverino, and M. Marzband, "Two-stage stochastic-based scheduling of multi-energy microgrids with electric and hydrogen vehicles charging stations, considering transactions through pool market and bilateral contracts," *International Journal of Hydrogen Energy,* 2023.

[11] A. Jani and S. Jadid, "Two-stage energy scheduling framework for multi-microgrid system in market environment," *Applied Energy,* vol. 336, p. 120683, 2023.

[12] W. Li, *Risk assessment of power systems: models, methods, and applications*: John Wiley & Sons, 2014.

[13] J. L. Lopez-Prado, J. I. Vélez, and G. A. Garcia-Llinas, "Reliability evaluation in distribution networks with microgrids: Review and classification of the literature," *Energies,* vol. 13, p. 6189, 2020.

[14] J. H. Carvalho, U. B. Schwartz, and C. L. Borges, "Copula based model for representation of hybrid power plants in non-sequential Monte Carlo reliability evaluation," *Sustainable Energy, Grids and Networks,* p. 101077, 2023.

[15] U. Agarwal, N. Jain, and M. Kumawat, "Reliability Analysis of Distribution System with Integration of Distributed Generation Resources," in *Optimal Planning and Operation of Distributed Energy Resources*, ed: Springer, 2023, pp. 235-258.

[16] R. Billinton and Z. Pan, "Historic performance-based distribution system risk assessment," *IEEE transactions on power delivery,* vol. 19, pp. 1759-1765, 2004.

[17] Z. Jing, C. Yu, F. Xi, F. Wu, Z. Tao, and P. Yang, "Reliability analysis of distribution network operation based on short-term future big data technology," in *Journal of Physics: Conference Series*, 2020, p. 012027.

[18] X. Chen, J. Tang, and W. Li, "Probabilistic operational reliability of composite power systems considering multiple meteorological factors," *IEEE Transactions on Power Systems,* vol. 35, pp. 85-97, 2019.

[19] Z. Ruifeng, H. Shuqing, F. Yu, W. Zhuoyue, and W. Yi, "Research on situation evaluation and prediction method for distribution network," in *2020 4th International Conference on Power and Energy Engineering (ICPEE)*, 2020, pp. 7-12.

[20] X. Chen, J. Qiu, L. Reedman, and Z. Y. Dong, "A statistical risk assessment framework for distribution network resilience," *IEEE Transactions on Power Systems,* vol. 34, pp. 4773-4783, 2019.

[21] P. Dehghanian, B. Zhang, T. Dokic, and M. Kezunovic, "Predictive risk analytics for weather-resilient operation of electric power systems," *IEEE Transactions on Sustainable Energy,* vol. 10, pp. 3-15, 2018.

[22] M. Mahzarnia, M. P. Moghaddam, P. Siano, and M.-R. Haghifam, "A comprehensive assessment of power system resilience to a hurricane using a two-stage analytical approach incorporating risk-based index," *Sustainable Energy Technologies and Assessments,* vol. 42, p. 100831, 2020.

[23] D. N. Trakas, M. Panteli, N. D. Hatziargyriou, and P. Mancarella, "Spatial risk analysis of power systems resilience during extreme events," *Risk Analysis,* vol. 39, pp. 195-211, 2019.

[24] A. Abdolahi, J. Salehi, F. S. Gazijahani, and A. Safari, "Assessing the potential of merchant energy storage to maximize social welfare of renewable-based distribution networks considering risk analysis," *Electric Power Systems Research*, vol. 188, p. 106522, Nov. 2020.

[25] A. Abdollahi and S. C. Khavar, "Enhancing the flexibility of decentralized energy resources through bi-level optimization in intra-day regional markets," *Renewable Energy Focus*, vol. 56, p. 100778, Mar. 2026.

[26] Y. Wu, Z. Lin, C. Liu, T. Huang, Y. Chen, Y. Ru*, et al.*, "Resilience enhancement for urban distribution network via risk-based emergency response plan amendment for ice disasters," *International Journal of Electrical Power & Energy Systems,* vol. 141, p. 108183, 2022.

[27] G. S. S. Gharehveran, K. Shirini, S. C. Khavar, and A. Abdollahi, "Optimizing day-ahead power scheduling: A novel MIQCP approach for enhanced SCUC with renewable integration," *e-Prime - Advances in Electrical Engineering, Electronics and Energy*, vol., p. 101022, May 19, 2025.

[28] W. A. Oraibi, B. Mohammadi-Ivatloo, S. Hosseini, and M. Abapour, "A resilience-oriented optimal planning of energy storage systems in high renewable energy penetrated systems," *Journal of Energy Storage,* vol. 67, p. 107500, 2023.

[29] M. Diahovchenko, G. Kandaperumal, and A. K. Srivastava, "Enabling resiliency using microgrids with dynamic boundaries," *Electric Power Systems Research,* vol. 221, p. 109460, 2023.

[30] Q. Shi, H. Wan, W. Liu, H. Han, Z. Wang, and F. Li, "Preventive allocation and post-disaster cooperative dispatch of emergency mobile resources for improved distribution system resilience," *International Journal of Electrical Power & Energy Systems,* vol. 152, p. 109238, 2023.

[31] M. Z. Gargari, M. T. Hagh, and S. G. Zadeh, "Preventive scheduling of a multi-energy microgrid with mobile energy storage to enhance the resiliency of the system," *Energy,* vol. 263, p. 125597, 2023.

[32] M. Rajabzadeh and M. Kalantar, "Improving the resilience of distribution network in coming across seismic damage using mobile battery energy storage system," *Journal of Energy Storage,* vol. 52, p. 104891, 2022.

[33] Y. Wang, A. O. Rousis, and G. Strbac, "Resilience-driven optimal sizing and pre-positioning of mobile energy storage systems in decentralized networked microgrids," *Applied Energy,* vol. 305, p. 117921, 2022.

[34] S. Samadi Gharehveran, K. Shirini, and A. Abdolahi, 'Optimizing Energy Storage Solutions for Grid Resilience: A Comprehensive Overview', Energy Storage Devices - A Comprehensive Overview. IntechOpen, Jan. 14, 2025. doi: 10.5772/intechopen.1006499.

[35] M. Nazemi, P. Dehghanian, X. Lu, and C. Chen, "Uncertainty-aware deployment of mobile energy storage systems for distribution grid resilience," *IEEE Transactions on Smart Grid,* vol. 12, pp. 3200-3214, 2021.

[36] A. Abdolahi and N. T. Kalantari, "State estimation of asymmetrical distribution networks by μ-PMU allocation: A novel stochastic two-stage programming," *Electric Power Systems Research*, vol. 213, p. 108738, 2022.